\documentstyle[12pt,twoside,psfig]{article}
\begin{document}
\title{\Large\bf Chameleon field and the 
                     late time acceleration \\of the universe}
\author{Narayan Banerjee{\footnote {E-mail: narayan@juphys.ernet.in}}~,
	Sudipta Das{\footnote{E-mail: sudipta\_123@yahoo.com}}~ and 
        Koyel Ganguly{\footnote{E-mail: koyel\_g\_m@yahoo.co.in}}\\ 
\it Relativity and Cosmology Research Centre,\\ 
\it Department of Physics, Jadavpur University,\\
\it Kolkata - 700032,\\
\it India.}
\date{}
\maketitle
\vspace{0.5cm}
\vspace{0.5cm}

\pagestyle{myheadings}
\newcommand{\be}{\begin{equation}}
\newcommand{\ee}{\end{equation}}
\newcommand{\bea}{\begin{eqnarray}}
\newcommand{\eea}{\end{eqnarray}}

\begin{abstract}
In the present work, it is shown that a chameleon scalar field having a 
nonminimal coupling with dark matter can give rise to a smooth transition 
from a decelerated to an accelerated phase of expansion for the universe. 
It is surprising to note that the coupling with the chameleon scalar 
field hardly affects the evolution of the dark matter sector, which still 
redshifts as $a^{-3}$.
\end{abstract}

\section{Introduction}
In the absence of a clear verdict in favour of any `dark energy' candidate, 
which drives the alleged late time acceleration of the universe \cite{1}, 
every possibilities are being thoroughly investigated so as to find out what 
can really negotiate this latest challenge that theoretical physics is 
exposed to. The accelerated expansion is counter-intuitive, as the dominating 
interaction in the dynamics of the universe is gravity which is  
attractive. So the `dark energy' sector has to produce an effective 
anti-gravity effect. However, the actual problem is even more stringent, both 
observations and theoretical requirements demand that the acceleration must 
have set in quite late in the evolution of the universe \cite{2}. The models 
are also required to be consistent with other observational data like those 
on the cosmic microwave background radiation. There are already some excellent 
reviews which summarize the problem and the efforts towards finding a 
consistent solution \cite{3}. Amongst the most talked about models, the 
quintessence models enjoy a fair deal of popularity. These models contain a 
scalar field endowed with a potential, so that the contribution to the 
pressure sector, $p_{\phi}=\frac{1}{2}{\dot{\phi}}^2 - V(\phi)$, can evolve 
to attain an adequately large negative value and drive the observed 
accelerated expansion. Given a particular temporal behaviour of 
the scale factor of the universe, it is always possible to find the requisite 
potential \cite{4}. Naturally there is a long list of potentials that 
can do the trick \cite{5}. None of them however can boast of a proper 
theoretical support from field theory.
\par In these scalar field models, the cold dark matter and dark energy are 
normally allowed to evolve independently. However, there are 
attempts to include an 
interaction amongst them so that one grows at the expense of the other ( see 
for example \cite{6} ). Nonminimally coupled scalar fields, where there is a 
coupling between the scalar field and geometry, also posed themselves as 
possible candidates for explaining the present acceleration. For example, 
Brans - Dicke theory cropped up as a possible arena \cite{7}. 
         \par A modification of the Brans - Dicke theory, where an interaction 
between the scalar field and the dark matter could be allowed, showed that 
the matter dominated era can have a transition from a decelerated to an 
accelerated expansion without the requirement of any dark energy \cite{8}.
 On the other hand, a dark energy interacting with Brans - Dicke scalar field 
can give rise to a late time acceleration for a very wide range 
of potentials \cite{9}.
\par A completely different approach has also been explored, where, 
unlike Brans - Dicke theory, the scalar field is allowed to have a 
``non-minimal'' coupling with the dark matter sector,
( but not with the Ricci scalar ) through an interference term in the 
action \cite{10}. This kind of a scalar field is dubbed as 
the ``chameleon field''. This field also proved quite useful in 
modelling an accelerated expansion of the universe. Many interesting 
cosmological possibilities with this chameleon field have been 
recently pointed out by Das, Corasaniti and Khoury \cite{11}. A recent 
review indicates the other gravitational implications of such 
a field \cite{12}.
\par In the present work, it has been shown that a simple 
chameleon field can indeed 
make room for a decelerated expansion in the early matter dominated 
era and allow for an accelerated expansion at the present epoch. 
This transition can take place quite smoothly at a recent past.
\par An intriguing feature of this field, as will be shown later, is 
that in spite of the interaction with the chameleon field the dark matter 
density falls off as $\frac{1}{a^3}$, where $a$ is the scale factor of 
the universe, exactly similar to the behaviour expected where the dark 
matter sector does not interact with dark energy and satisfies its 
conservation equation by itself. So the apprehension that there could 
be discrepancy between a chameleon model and the expected $(1+z)^3$ 
dependence in the clustering of the cold dark matter \cite{11} is ruled 
out.  

\section{Field Equations and Results :~}
The relevant action is 
\be\label{action}
A = \int\left[\frac{R}{16 \pi G} + \frac{1}{2}{\phi}_{,\mu}{\phi}^{,\mu} 
                 -V(\phi) + f(\phi) L_{m}\right]{\sqrt{-g}}~d^4x~, 
\ee 
where $R$ is the Ricci scalar, $G$ is the Newtonian constant of gravity and 
$\phi$ is the chameleon scalar field with a potential $V(\phi)$. 
Unlike the usual 
Einstein - Hilbert action, the matter Lagrangian $L_{m}$ is modified as 
$f(\phi)L_{m}$, where $f(\phi)$ is an analytic function of $\phi$. This term 
brings about the nonminimal interaction between the cold dark matter and 
chameleon field. It deserves mention that a string inspired dilaton field 
also has a similar coupling with the matter sector \cite{13}. For a spatially 
flat FRW metric, the line element is given by 
\be 
ds^2 = dt^2 - a^2(t) \left[ dr^2 + r^2d{\theta}^2 + r^2\sin^2{\theta} 
                      d{\phi}^2 \right]~.
\ee
Variation of action (\ref{action}) with respect to the metric tensor components 
yields the field equations as 
\be\label{fe1}
3\frac{{\dot{a}}^2}{a^2} = \rho_{m}f + \frac{1}{2}{\dot{\phi}}^2 + V(\phi)~,
\ee
\be\label{fe2}
2\frac{\ddot{a}}{a} + \frac{{\dot{a}}^2}{a^2} = -\frac{1}{2}{\dot{\phi}}^2 
                                                   + V(\phi),~
\ee
where $8 \pi G = 1$. As we are interested in a matter dominated universe, the 
fluid is taken in the form of pressureless dust ( $p_{m} = 0$ ); $\rho_{m}$ 
stands for the contribution from the cold dark matter to the energy density 
and an overhead dot indicates differentiation with respect to the cosmic 
time $t$. 
\par Also, variation of the action (\ref{action}) with respect to $\phi$ 
provides the wave equation for the chameleon field as 
\be\label{wave}
\ddot{\phi} + 3 H \dot{\phi} = - V' -\rho_{m} f' 
\ee
where a prime indicates differentiation with respect to $\phi$. 
\\
From equations (\ref{fe1}), (\ref{fe2}) and (\ref{wave}), one can easily 
arrive at a relation
\be
\left(\rho_{m}f\right)^{.} + 3H\rho_{m}f = \rho_{m}\dot{\phi}f'
\ee
which readily integrates to yield 
\be\label{matter}
\rho_{m} = \frac{\rho_{0}}{a^3}~,
\ee
$\rho_{0}$ being a constant.
\\
So it is quite clear, as indicated before, the cold dark matter (CDM) indeed 
redshifts as $(1 + z)^3$. The coupling factor $f(\phi)$ in equation 
(\ref{fe1}) can only tune the epoch at which the baton of dominance shifts 
from the CDM to the dark energy sector.
Now, out of equations (\ref{fe1}), (\ref{fe2}), (\ref{wave}) and (\ref{matter}),
 only three are independent equations as the fourth one can be derived
 from the other three by using Bianchi identity. So one is left with four
unknowns $a$, $\phi$, $V(\phi)$, $f(\phi)$ and three independent equations.
So we make an ansatz\\
\be\label{hubble}
H = \frac{\dot{a}}{a} = e^{(1-{\gamma}a^2)/{\alpha a}}
\ee
where $\gamma$ and $\alpha$ are positive constants.\\
Obviously this choice is made to track the observed expansion of the
 universe. The system of equations is now closed, but the high degree of
 non linearity makes it difficult to get a set of solutions. In order to
 present a complete model, we make the simplifying assumption\\
\be\label{potential}
V = \beta{\dot{\phi}^2}~,
\ee
$\beta$ being a positive constant. This assumption of the kinetic 
and the potential part of the chameleon field being proportional to
 each other is indeed restrictive. But the field equations are satisfied.
The constant $\beta$ must be greater than $\frac{1}{2}$, so that the 
effective pressure is negative (equation (\ref{fe2})). With the assumption
 (\ref{potential}), the field equations are satisfied with
\be
{\rho}_mf=\frac{H^2}{(2\beta-1)}\left[(4\beta+2)\left(\frac{1}{{\alpha}a}
+\frac{{\gamma}a}{\alpha}\right)-6\right]~,
\ee
\be
{\dot{\phi}}^2 = \frac{H^2}{(2\beta-1)}\left[6-4\left(\frac{1}{{\alpha}a}
+\frac{{\gamma}a}{\alpha}\right)\right]
\ee
and
\be
f = \frac{2a^3H^2}{{\rho}_0(2\beta-1)}
\left[(2\beta+1)\left(\frac{1}{{\alpha}a}+\frac{{\gamma}a}{\alpha}\right)
              -3\right]~.
\ee
\\
With $H$ given by the equation(\ref{hubble}), all the relevant quantities 
are now known in terms of the scale factor $a$, and the evolution of each 
quantity can be found out.

\par
\begin{figure}[!h]
\centerline{\psfig{figure=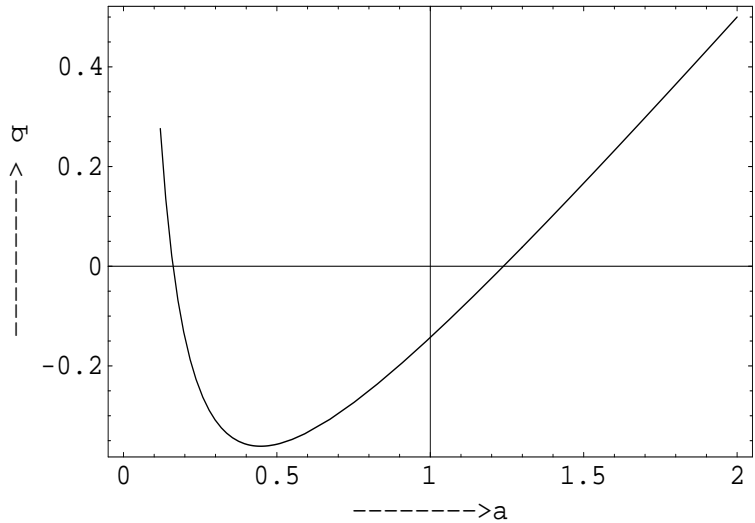,height=70mm,width=100mm}}
\caption{\normalsize{Plot of $q$ vs. $a$ for $\alpha = 7$ and $\gamma = 5$.}}
\label{fig1}
\end{figure}

 Knowing that the ansatz (\ref{hubble}) and (\ref{potential}) can
yield a consistent set of solutions, one can now talk about the
 deceleration parameter $q$, which can be written
 ( from equation (\ref{hubble}) ) as 
\be\label{dec}
q=\frac{1+{\gamma}a^2}{{\alpha}a}-1~.
\ee 
Equation (\ref{dec}) indicates that one can obtain a negative $q$ at
present ( $a = 1$ ) if $\alpha > 1 + \gamma$. We work out the model for
 $\alpha = 7$ and $\gamma = 5$ which satisfies this condition.
 A plot of $q$ vs $a$ shows that $q$ enters into a negative value
regime from a positive value and one has the desired feature of
negative $q$ at $a = 1$, i.e, present epoch. This indicates that
 the universe has undergone a transition from a decelerated phase of
expansion to an accelerated one in the recent past in the matter
dominated era itself as expected both theoretically \cite{2} and
observationally \cite{14}. From figure \ref{fig1} it is seen that this 
flip in $q$ takes place at around $a\approx0.17$ which corresponds 
to $z \approx 4.9$ which is quite reasonable as argued by Amendola \cite{15}.

\par The additional feature of this model is that $q$ has another
 signature flip from a negative to positive direction at around
 $a\approx1.25$ which indicates that the universe re-enters a
 decelerated phase of expansion in `future' and thus `phantom menace'
can be avoided - the universe does not have a singularity of infinite
volume and infinite rate of expansion in a `finite' future. Also the
 nature of the plot is not crucially sensitive to values of $\alpha$
and $\gamma$ chosen; the time when flip occurs shifts a little with
different choices of $\alpha$ and $\gamma$. 
\par
\begin{figure}[!h]
\centerline{\psfig{figure=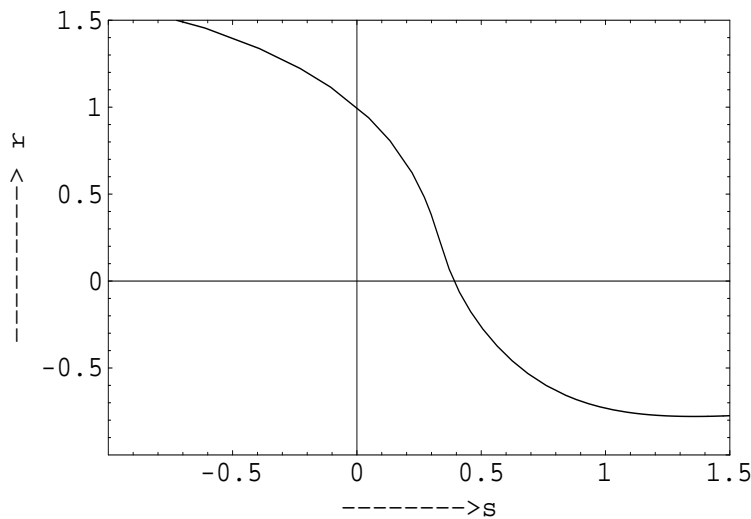,height=70mm,width=100mm}}
\caption{\normalsize{Plot of $r$ as a function of $s$ for $\alpha = 7$ and 
              $\gamma = 5$.}}
\label{fig2}
\end{figure}

 The statefinder parameters $\{r,s\}$, introduced by Sahni
 et al \cite{16}, has also been computed for this model. These parameters
can effectively differentiate between different forms of dark energy and
provide a simple diagnostic regarding whether a particular model fits 
into the basic observational data. These parameters are defined as  
\be
r = \frac{\stackrel{...}{a}}{a H^3} ~~~~\mathrm{and}~~~~ 
      s=\frac{r-1}{3(q-1/2)}~.
\ee  
As present observations facilitate the study of the evolution of the 
deceleration parameter $q$, these statefinder parameters have become useful
as they involve third order derivative of the scale factor `$a$'.\\
For the present model, the statefinder pair comes out as
\be
r=-2\left[\frac{2{\gamma}a}{\alpha}+\frac{1}{{\alpha}a}\right]+
2{\left[\frac{1}{{\alpha}a}+\frac{{\gamma}a}{\alpha}\right]}^2+1
\ee
and
\be
s=\frac{4({\gamma}^2{a}^4-2{\gamma}{\alpha}{a}^3+2{\gamma}{a}^2-{\alpha}a+1)}
{3(2{\gamma}{\alpha}{a}^3-3{\alpha}^2{a}^2+2{\alpha}a)}~.
\ee
We now plot $r(s)$ (figure \ref{fig2}) for the same values of 
$\alpha$ and $\gamma$. The plot shows that this model does not mimic the 
most talked about models like the $\Lambda$CDM or quintessence. So the 
model has distinguishing features, and hence open to the judgement regarding 
the suitability compared to the other competing models. 
\section{Discussion :}
The smooth transition from a decelerated to an accelerated phase of expansion 
driven by the chameleon field makes the latter worthy of attention. The model 
presented here is restricted, but it clearly shows that the nonminimal 
coupling of the field with the CDM sector governs the onset of the 
acceleration, but it hardly affects the $a^{-3}$ behaviour of the CDM. This 
field thus warrants more attention. If one could find a coupling $f(\phi)$ 
so that the chameleon field has an oscillatory behaviour ( with a small 
amplitude ) at the beginning of the matter dominated epoch, but grows later to 
dictate the dynamics of the universe, it would solve many a fine tuning 
problems.
\section{Acknowledgement}
Authors SD and KG wish to thank CSIR and BRNS(DAE) respectively 
for financial support.

\end{document}